\documentclass[pre,a4paper,oneside,nofootinbib]{revtex4}

\usepackage{color} 
\usepackage[normalem]{ulem} %
\usepackage{graphicx}
\usepackage{hyperref}
\usepackage{amsmath}
\usepackage{amssymb}

\usepackage{graphicx} 

\usepackage{amsmath}
\usepackage{xcolor}

\usepackage[normalem]{ulem} 

\usepackage{xcolor}
\usepackage{lipsum}

\usepackage{hyperref}

\begin{document}

\title{How buildings change the fundamental allometry}

\author{Fabiano L.\ Ribeiro [0000-0002-2719-6061]}
\email{fribeiro@ufla.br}
\affiliation{Department of Physics (DFI), Federal University of Lavras (UFLA), Lavras MG, Brazil}

\author{Peiran Zhang}
\email{peiranzhang1999@gmail.com}
\affiliation{School of Systems Science Beijing Jiaotong University Beijing}
\affiliation{Leibniz Institute of Ecological Urban and Regional Development (IOER), Dresden, Germany}

\author{Liang Gao}
\email{lianggao@bjtu.edu.cn}
\affiliation{School of Systems Science Beijing Jiaotong University Beijing}

\author{Diego Rybski [0000-0001-6125-7705]}
\email{ca-dr@rybski.de}
\affiliation{Leibniz Institute of Ecological Urban and Regional Development (IOER), Dresden, Germany}
\affiliation{Complexity Science Hub Vienna, Josefst\"adterstrasse 39, A-1090 Vienna, Austria}


\date{\today}


\begin{abstract}
We demonstrate that the original fundamental allometry alone cannot accurately describe the relationship between urban area and population size.
Instead, building height is a third factor that interplays with area and population. 
To illustrate this, we propose a straightforward model based on the idea that \emph{city area is the result of people's desire to live close to one another while also having sufficient living space}.
This leads to a more general form of fundamental allometry (relating area, population, \emph{and} building height).
Our argument is supported by empirical data from different countries. 
\end{abstract}

\maketitle

\section*{Introduction}

\emph{Urban scaling} refers to an urban indicator that is related to city size in terms of population, and in many cases, non-linear power-law association is reported \cite{BettencourtLHKW2007}.
In a recent review, mathematical models explaining urban scaling have been presented and compared \cite{RibeiroR2023}.
A special form of urban scaling is the relation between city area $A$ and city population $N$. 
We call it \emph{fundamental allometry} because both quantities represent a measure of size, and relating population and area implicitly brings density into play, which itself is an important quantity.
In the form 
\begin{equation}\label{Eq_fund}
A\sim N^\beta \, ,    
\end{equation}
most studies report $\beta<1$ \cite{BattyF2011,BurgerOHWSLBCFHHKEE2022}, implying higher population density in large cities.
This finding is remarkable and even obtained for ancient urban systems and societies \cite{OrtmanCSB2014,HamiltonMWB2007,LoboBSO2019}.
For a city resembling a hemisphere, assuming the population is proportional to the volume ($N\sim r^3$) and assuming the area of a circle ($A\sim r^2$), one gets $A\sim N^{2/3}$ \cite{NordbeckS1971}.
Certainly, cities are not hemispherical, and in particular, the vertical dimension has a very different character compared to the horizontal ones.
Consequently, more elaborated derivations for the scaling exponent have been proposed \cite{BettencourtLMA2013,LoufB2014,Deppman2024}. 
Here, we present a simple derivation that, as we will see below, leads to the same results as more complex models.
In addition, our results suggest that the average building height represents a third quantity that interplays with area and population.

\section*{The model}
We begin by considering that the fundamental allometry emerges from an equilibrium area $A^*(N)$;  i.e., \ for a population count $N$, there is an equilibrium area (indicated by the asterisk) and vice versa.
Based on the following considerations, we argue that this equilibrium results from two forces.

\begin{enumerate}
\item  Urban dwellers want to be close to others and to facilities, infrastructure, etc.
According to this ``\emph{force}'', a given population wants to concentrate on a small city area, implying high density.
In economics, this attraction would be a ``\emph{centripetal force}''. 
Here, we will refer to this force simply by ``\emph{atractive force}'';

\item  Urban dwellers want to inhabit houses and apartments with large individual living areas.
According to this second ``force'', a given population spreads out in a large city area, implying low density.
In economics, this repulsion would be a ``\emph{centrifugal force}''.
 We will refer to this force simply by ``\emph{repulsive force}''. 
\end{enumerate}

Analogous to physics, this system of forces can be described by a \textit{potential function} $U(A)$, and then the resulting force, representing the intensity with which the system attempts to restore equilibrium, can be expressed as $F = -\frac{dU}{dA}$.
Using this idea, we can consider that a city will be in equilibrium when both forces add up to zero, which is the case where the potential exhibits a minimum, i.e., \ when $F(A^*) = -\frac{dU(A)}{dA} = 0$ (necessary condition).
We must find a mathematical expression for the potential $U(A)$ to describe the system formally.
Involving the two forces, the potential must be the result of two other potentials, i.e., \ $U(A)=u_1(A)+u_2(A)$. 
The first, $u_1(A)$, describes the attractive force and must be an increase function of the area, in the sense that its minimum will be when the area is small. 
The second, $u_2(A)$, describes the repulsive force 
and must be a decreased function of the area because its minimum will be when the area is large. 
This idea is illustrated in Fig.~\ref{fig:UxA}.

\subsection*{Attractive force: average distance between the individuals}
Accounting for the urge of urban dwellers to be close to one another, we can use the average distance $\bar{d}$ between locations/individuals in the city, 
calculated as

\begin{equation}
    \bar{d} = \frac{2}{N(N-1)} \sum_{i=1}^N \sum_{j \ne i} r_{ij} 
    \label{eq:doublesum}
    \, .
\end{equation}
Here, \( r_{ij} \) represents the distance between individuals \( i \) and \( j \) (more precisely, the distance between their residences).
The respective potential $u_1(A)$ must increase with $\bar{d}$ while its minimum happens as $\bar{d}$ approaches zero, reflecting that dwellers would want to live as close as possible to each other. 
It justifies considering $u_1 \sim \bar{d}$.

In addition, for a fixed population, the average distance increases with the area.
In the recent work \cite{RibeiroLBR2024}, 
it is shown that when a set of points is distributed in 2D space, forming a macroscopic object with a fractal dimension $D_f$ and area $A$, then the average distance between the points scales as $\bar{d} \sim A^{1/D_f}$ (the solution of the double sum in Eq.~\ref{eq:doublesum}).
This is precisely the context here, and therefore, it is reasonable to consider that $u_1(A) \sim \bar{d} \sim A^{1/D_f}$.

\subsection*{Repulsive force: individual living area}

For the second potential, $u_2(A)$, we can define the \textit{average floor area per capita}, say $\bar{a}$, which can be calculated by the total floor area of the city divided by the population.
That is, if 
$\bar{h}$ represents the \textit{average building height}, e.g., measured in number of floors, then 

\begin{equation}
   \bar{a} \sim \frac{\bar{h}A}{N}
    \, .
\end{equation}
Here, the total floor area $\bar{h}A$ can also be interpreted as the \textit{average building volume}.   
In the particular case where $\bar{h}=1$ floor, the average floor area per capita is the inverse of the population density (or area per capita), i.e., $\bar{a} \sim A/N$.
The respective potential must decrease with $\bar{a}$ since its minimization occurs only when $\bar{a}$ approaches infinity.
This indicates that dwellers desire as much individual space as possible, and then it is reasonable to consider $u_2 \sim 1/\bar{a}$, fulfilling the requirements.

\begin{figure}
\centering
\includegraphics[width=0.7\linewidth]{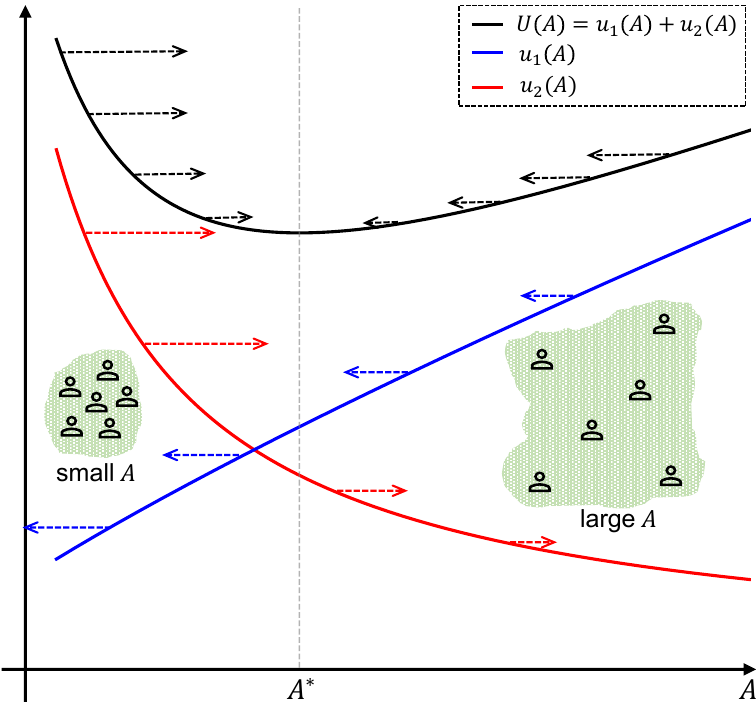}
\caption{\label{fig:UxA} 
The potential function $U(A)$ -  curve in black -  as a combination of two other potentials by the form $U(A) = u_1(A) + u_2(A)$, for a fixed population $N$.
The potential $u_1(A)$ - curve in blue -  is responsible for an attractive force (the blue arrows) that represents the desire of the people to live close to each other -- the reason for the cities to exist.
The potential $u_2(A)$ - curve in red - is responsible for a repulsive force (the red arrows) that represents the desire of the people to live in large housing/living areas.
The minimum of $U(A)$ (black curve) determines the city's equilibrium area $A^*$.
The arrows in black, obtained by the equation $F = -\frac{dU}{dA}$, represent the intensity and direction (the force) driving the system back to equilibrium.
}
\end{figure}

\section*{Results}

Combining both functions, we get the potential function
\begin{equation}
    U(A) = A^{1/D_f} + \frac{N}{\bar{h} A}
    \, .
\end{equation}
This function $U(A)$ depends on the area $A$, and its minimum corresponds to the equilibrium area $A^* $ (see Fig~\ref{fig:UxA}), which we obtain from $dU/dA = 0$, yielding
\begin{equation}
    A^* \sim \left( \frac{N}{\bar{h}} \right)^{\frac{D_f}{D_f +1}} \quad 
    \text{ and } A^* \sim N^{\frac{D_f}{D_f +1}} \text{ for constant }\bar{h}
    \label{eq:optimalA}
    \, .
\end{equation}
This result allows three observations.

First, since the fractal dimension is positive, the scaling exponent is smaller than one, $D_f/(D_f+1)<1$, consistent with the literature. 
Second, it is equivalent to the expression by Bettencourt~2013 \cite{BettencourtLMA2013} 
who -- from \emph{different} arguments\footnote{
Bettencourt derives the fundamental allometry scaling exponent by considering that the probability of encounters between people decreases with the area. Additionally, he posits that salaries, which depend on the probability of these encounters, must ensure individuals have access to the entire city.}
-- derives the \emph{identical} exponent $\beta=\frac{D_f}{D_f+1}$.
Formally, Eq.~(\ref{eq:optimalA}) is also compatible with the result by Louf \& Barthelemy 
\cite{LoufB2014} who obtain $\beta=\frac{2\mu}{2\mu+1}$, where $\mu$ is an exponent reflecting the resilience of the transport network to congestion \cite{RibeiroR2023}.

\begin{figure}
\centering
a)\includegraphics[width=0.3\linewidth]{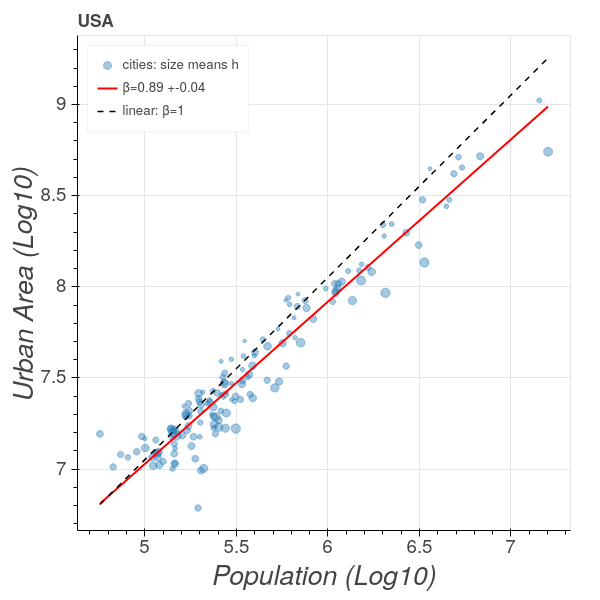}
b)\includegraphics[width=0.3\linewidth]{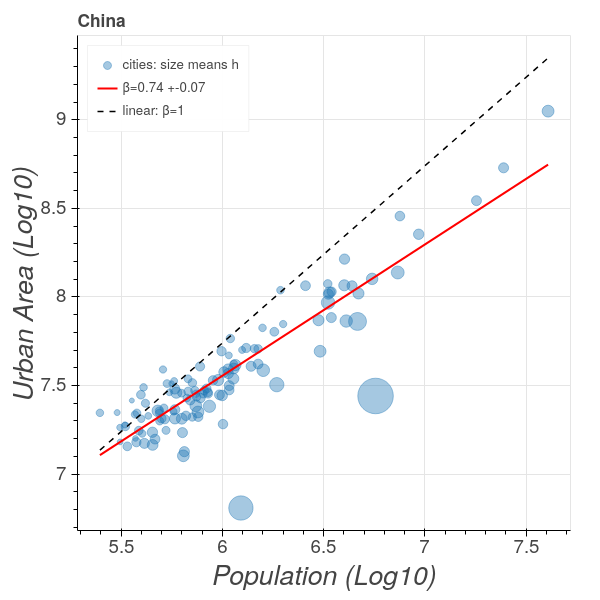}
c)\includegraphics[width=0.3\linewidth]{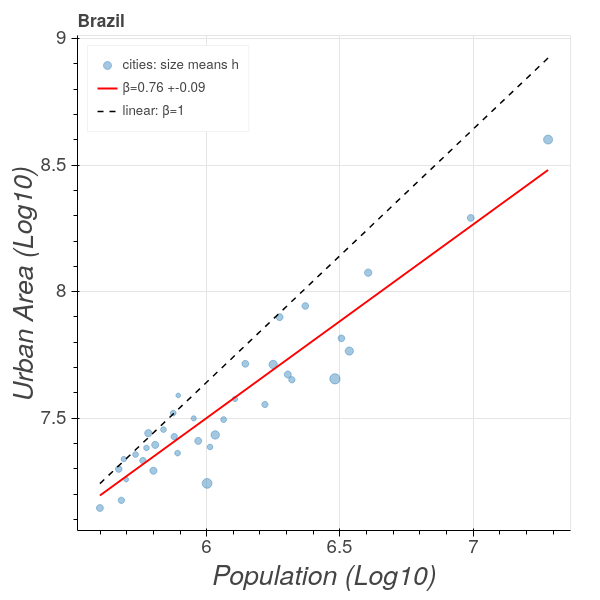}
d)\includegraphics[width=0.3\linewidth]{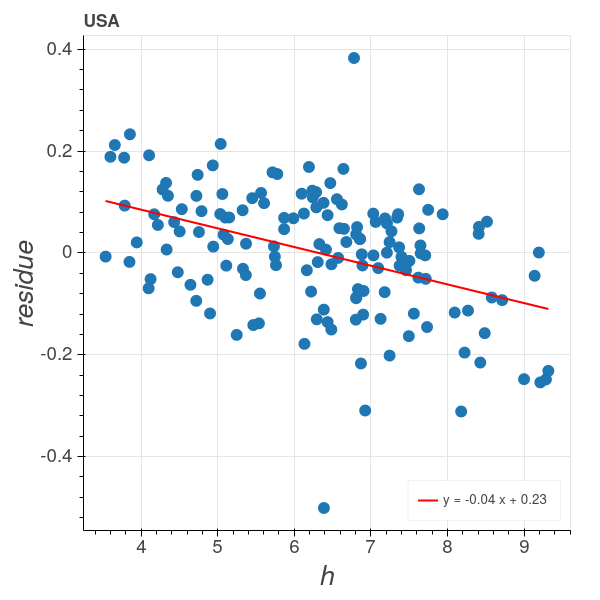}
e)\includegraphics[width=0.3\linewidth]{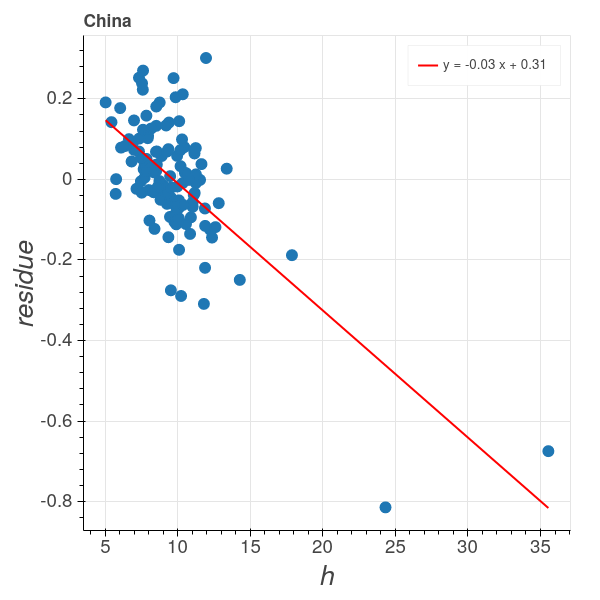}
f)\includegraphics[width=0.3\linewidth]{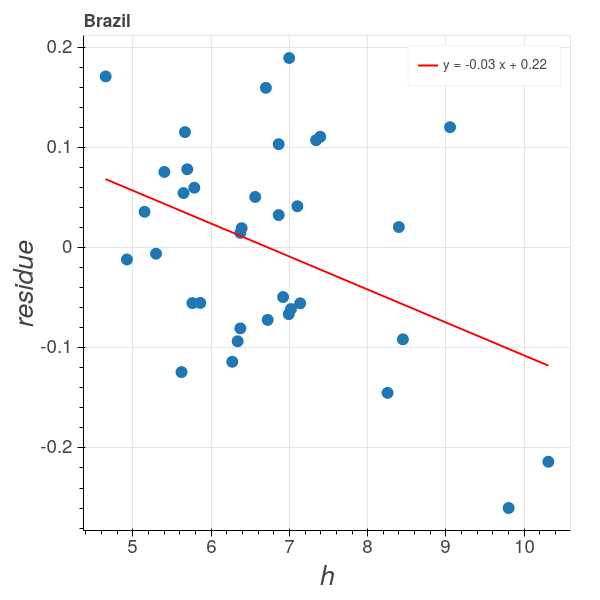}
g)\includegraphics[width=0.32\linewidth]{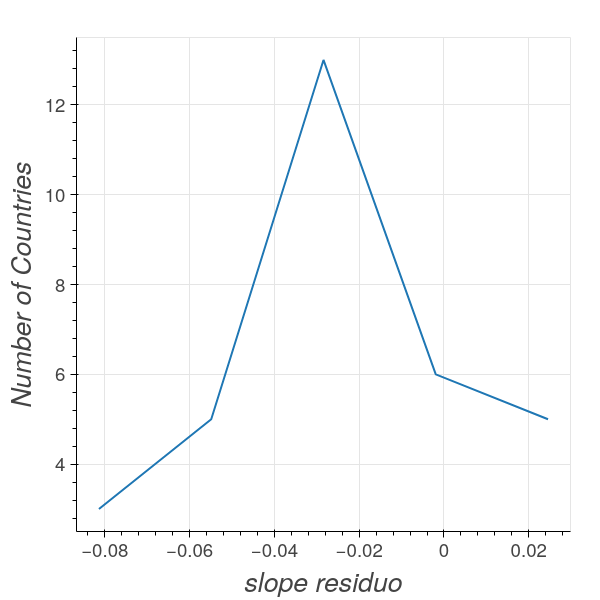}
\caption{\label{fig:resuduals} 
Three Figures on top (a,b,c): Urban Area as a Function of Population Size for Three Continental-Sized Countries: USA, China, and Brazil. 
Each dot represents one city (UCDB), with the size of the dots indicating the city's average building height. The red line corresponds to the fit of the original fundamental allometry (i.e., \ $A \sim N^{\beta}$). Visually, it is apparent that cities with larger building heights tend to fall below the red line. 
To systematically justify this observation, we examine the residuals of the cities as a function of average building height, as presented in the three figures in the middle (d,e,f).
These residuals, which measure the deviation from the predicted fundamental allometric relationship, exhibit a negative slope in relation to average building height. This pattern holds not only for these three countries but also for the majority of the 32 countries included in the study, as shown in the histogram below (g). The negative slope of the residuals indicates that cities with taller average building heights tend to fall below the line predicted by the original fundamental allometric relationship.
Dataset: \emph{GHSL Global Human Settlement Layer}
\cite{ghsl}
and \emph{World Settlement Footprint} \cite{wsf}.}
\end{figure}

Third, Eq.~(\ref{eq:optimalA}) indicates that not only the population size is necessary to infer the area of the city, as the previous models suggest, but also the building height.
In Fig.~\ref{fig:resuduals}, we examine how the building height affects the residuals (which are also known as \emph{Adjusted Metropolitan Indicators} \cite{BettencourtLSW2010}).  
In this figure, the fundamental allometry is displayed, where each dot represents one city based on 
the UCDB City Identification Standards, and the size of the dots indicates the average building height.
One can see that cities consisting of taller buildings tend to be located below the regression line, and cities with smaller average building heights are above -- supporting Eq.~(\ref{eq:optimalA}).
The analysis presented in this figure shows that the residuals typically decrease (and become negative) with increasing average building height. 

In this sense, if we redefine the residuals as $R=\ln{(c(N/\bar{h})^\beta)}-\ln{(A_0N^\beta)}$ (difference between the prediction of Eq.~(\ref{eq:optimalA}) and the original fundamental allometry Eq.~(\ref{Eq_fund})), where $c$ and $A_0$ are constants, then $\bar{h}=(A_0/c)^{-1/\beta}\text{e}^{-R/\beta}$.
Expanding the exponential function for small $R$ we obtain 
$R = -a \cdot \bar{h} + b$, where $a$ and $b$ are constants that depend on $\beta$. 
That is, a decreasing linear function between residual and average building height, as in 
the middle panels of Fig.~\ref{fig:resuduals}.
This result aligns with intuition -- smaller city areas require taller buildings to accommodate the same number of urban dwellers.

A particularly interesting case is the cities of Hong Kong and Sha Tin, both in China. When considering only the original fundamental allometry, these cities (UCDB) appear as outliers, positioned significantly below the fundamental allometry fit curve (see figure~\ref{fig:resuduals} - b and e). This is because they have large populations within relatively small areas, which contradicts the simple expectations of the original fundamental allometry. However, their higher average building heights explain this discrepancy, allowing them to accommodate more people in smaller areas. This situation aligns with the results of the approach we are discussing here.

\section*{Final remarks}
In conclusion,  the vertical growth of the cities must be considered when relating area and population size.
Not taking the building height variability into account increases the noise of the fundamental allometry (or reduces the correlations).
Thus, the simple fundamental allometry Eq.~(\ref{Eq_fund}) is just a kind of first-order approximation.
Understanding the complexity of urban scaling, including the role of building height, as introduced in \cite{SchlapferLB2015,MolineroT2021}, represents an ongoing challenge.

It is interesting to observe that if we take above's argument $A\sim r^2$, $N\sim r^3$, and introduce the fractal dimension (instead of $D=2$), i.e.\ $A\sim r^{D_f}$ and $N\sim r^{D_f+1}$, then we get the same $\beta=\frac{D_f}{D_f+1}$ only that the vertical dimension is not fractal.
More research is necessary to understand and describe the different characteristics of vertical and horizontal dimensions in cities.
The equilibrium model we present here is very simple, and one can think of more complete ones, possibly distinguishing and addressing vertical and horizontal dimensions.

\vskip 2 cm

{\bf Acknowledgement:} F.\ L.\ Ribeiro thanks CNPq (grant numbers 403139/2021-0 and 424686/2021-0) and Fapemig (grant number APQ-00829-21) for financial support. P.Z. and L.G. are supported by the National Natural Science Foundation of China (72288101 and 72242102). P.Z. acknowledges support from the program of China Scholarship Council (No. 202307090047).
D.\ Rybski acknowledges financial support from German Research Foundation (DFG) for the projects UPon (\#451083179) and Gropius (\#511568027).

\end{document}